\documentclass[12pt]{article}
\usepackage{amsfonts,amsmath,amssymb}
\usepackage{color}
\def\C{\mathbb{C}}

\def\bq{ \begin{equation} }
\def\eq{ \end{equation} }
\def\ben{ \begin{eqnarray} }
\def\en{ \end{eqnarray} }

\def\frac#1#2{{#1\over #2}}

\def\on#1#2{\mathop{\vbox{\ialign{##\crcr\noalign{\kern2pt}
$\scriptstyle{#2}$\crcr\noalign{\kern2pt\nointerlineskip}
\kern-2pt$\hfil\displaystyle{#1}\hfil$\crcr}}}\limits}

\def\ldb{\mathopen{\{\!\!\{}} \def\rdb{\mathclose{\}\!\!\}}}

\begin{document}

%%%%%%%%%%%%%%%%%%%%%%%%%%%%%%%%%%%%%%
\baselineskip=15pt
%\begin{flushright}
%Draft\\
%12/04/2003
%\end{flushright}
\vspace{1cm} \centerline{\LARGE  \textbf{Classification of constant solutions for
}}
\vskip0.4cm
\centerline{\LARGE  \textbf{associative Yang-Baxter equation on $gl(3)$ }}

\vskip1cm \hfill
\begin{minipage}{13.5cm}
\baselineskip=10pt
{\large \bf
   V Sokolov ${}^{1}$} \\ [2ex]
{\footnotesize
${}^{1}$ Landau Institute for Theoretical Physics, Moscow, Russia }\\

\vskip1cm{\bf Abstract}
\bigskip

We find all non-equivalent constant solutions for classical associative Yang-Baxter equation for $gl(3)$. New examples found in the classification yield the corresponding quadratic  trace Poisson brackets, double Poisson structures on free associative algebra with three generators and anti-Frobenius associative algebras.

\end{minipage}

\vglue1cm \textbf{Address}:
Landau Institute for Theoretical Physics, Kosygina 2, 119334, Moscow, Russia

\textbf{E-mail}:
sokolov@itp.ac.ru \newpage

\section{Introduction}
 Consider skew-symmetric solutions of the associative Yang-Baxter equation (another name is the Rota-Baxter equation) \cite{Rota,Sch}
\begin{equation} \label{yang}
r^{23}r^{12}+r^{31}r^{23}+r^{12}r^{31}=0, \qquad r^{12}=-r^{21}.
\end{equation}
Here  $r$ is a linear operator on $V\otimes V,$ where $V$ is $n$-dimensional vector space, all operators in (\ref{yang}) act in $V\otimes V\otimes V$,  and $~r^{ij}$ means the operator $~r$ acting in the
product of the $i$-th and $j$-th components. Let $e_\alpha,~\alpha=1,...,m$ be a basis in $V$. If
$$r(e_\alpha\otimes
e_\beta)=r^{\sigma \epsilon}_{\alpha \beta}e_\sigma\otimes e_\epsilon,
$$
then (\ref{yang}) is equivalent to the following system of algebraic equations:
\begin{equation}\label{r1}
r^{\sigma \epsilon}_{\alpha\beta}=-r^{\epsilon\sigma}_{\beta\alpha}.
\end{equation}
and
\begin{equation}\label{r2}
r^{\lambda\sigma}_{\alpha\beta}
r^{\mu\nu}_{\sigma\tau}+r^{\mu\sigma}_{\beta\tau} r^{\nu\lambda}_{\sigma\alpha}+r^{\nu\sigma}_{\tau\alpha} r^{\lambda\mu}_{\sigma\beta}=0.
\end{equation}
The coefficients of the tensor $r$  are
transformed under a change of the basis $e_{\alpha}$ in the standard way:
\begin{equation}\label{basis}
r_{\alpha\beta}^{\gamma \sigma} \to g^{\lambda}_\alpha g^{\mu}_\beta h_{\nu}^\gamma h_{\epsilon}^\sigma
\,r_{\lambda \mu}^{\nu \epsilon},
 \end{equation}
where $g_\alpha^\beta h_\beta^\gamma=\delta_\alpha^\gamma$.

It is easy to verify that the system of algebraic equations (\ref{r1}), (\ref{r2}) is invariant with respect to the involution
\begin{equation}\label{sig}
T: \quad  r^{\gamma \delta}_{\alpha\beta}\to r_{\gamma \delta}^{\alpha \beta}.
\end{equation}

In is known that any solution of (\ref{r1}), (\ref{r2}) gives rise to
\begin{itemize}
\item the trace quadratic Poisson bracket (see \cite{ MikSok, OdRubSok1}
\begin{equation}\label{Poisson}
\{x^j_{i,\alpha},x^{j^{\prime}}_{i^{\prime},\beta}\}=
r^{\gamma\epsilon}_{\alpha\beta}x^{j^{\prime}}_{i,\gamma}x^j_{i^{\prime},\epsilon},
\end{equation}
where $x^j_{i,\alpha}$ are entries of matrices $x_{\alpha},$ $\alpha=1,...,n$;
\item the double Poisson bracket \cite{VdB,OdRubSok2}
\begin{equation} \label{dquad}
\ldb x_{\alpha}, x_{\beta}\rdb =r_{\alpha \beta}^{u v} \, x_u \otimes x_v
 \end{equation}
 on the free associative algebra $A=\C<x_1,\ldots,x_m>$;
 \item an  {\it anti-Frobenius} associative subalgebra  in $Mat_n$   \cite{Agui,OdRubSok1}.
\end{itemize}

All known solutions are generated by the anti-Frobenius algebras of $n\times n$-matrices with $k$ zero rows (or columns), where $k$ is any divisor of $n$.

In the simplest case $n=2$ all solutions of (\ref{r1}), (\ref{r2}) up to transformations (\ref{basis}) and the scaling $r\to \lambda r$  were listed in \cite{Agui,OdRubSok2}. There exist the following two solutions:

{\bf Case 1.}  $r^{21}_{22}=-r^{12}_{22}=1$.

{\bf Case 2.} $r^{22}_{21}=-r^{22}_{12}=1$. \newline
We present here non-zero components of the tensor $r$ only. These solutions correspond to $2\times 2$-matrices with one zero row and one zero column, correspondingly.

Two solutions are called {\it equivalent} if they are related by transformations (\ref{basis})  or by the scaling $r\to \lambda r$. The main goal of this paper is a classification of all solutions (\ref{r1}), (\ref{r2}) in the case $n=3$ up to the equivalence.

 \section{Preliminary classification}

Because of the skew-symmetricity of $r$ we have 36 unknowns and several hundred equations in system (\ref{r1}). In addition, the action (\ref{basis}) of $gl(3)$  produces parameters in solutions. Therefore we have no chances to solve the system straightforwardly by computer algebra tools. However it turns to be possible if we use a sort of a deformation technique.

Any tensor $r$ after transformation (\ref{basis}) with the matrix $g=diag(1,1,\varepsilon)$ and multiplication by $\varepsilon^2$ has the form
\begin{equation}\label{deform}
r^{\alpha \beta}_{\gamma \delta}=a^{\alpha \beta}_{\gamma \delta}+\varepsilon b^{\alpha \beta}_{\gamma \delta}+\varepsilon^2 c^{\alpha \beta}_{\gamma \delta}+\varepsilon^3 d^{\alpha \beta}_{\gamma \delta}+\varepsilon^4 f^{\alpha \beta}_{\gamma \delta}.
\end{equation}
We denote the set of all components $a^{\alpha \beta}_{\gamma \delta}$ by $a$ and so on. It is easy to verify that the set $a$ may contain only one (up to the skew-symmetricity) non-zero component $a^{33}_{12}$. The set $b$  contains the following ten non-zero elements:
$$
b^{31}_{11},\,\,   b^{31}_{21},\,\, b^{31}_{12},\,\, b^{31}_{22},\,\, b^{32}_{11},\,\,  b^{32}_{21},\,\, b^{32}_{12},\,\, b^{32}_{22},\,\,  b^{33}_{31},\,\, b^{33}_{32}
$$
and skew-symmetric to them.
The number of non-zero independent elements in $c$, $d$ and $f$ equals 14, 10 and 1, correspondingly.

We are looking for solutions of (\ref{r1}), (\ref{r2}) in the form (\ref{deform}). It allows us to separate equations from (\ref{yang}) into several groups corresponding to different powers of $\varepsilon$ and to solve them more or less sequentially.

In this section we find a complete list of families of solutions for (\ref{yang}). We use here mostly transformations (\ref{basis}) with $g^1_3=g^2_3=g^3_1=g^3_2=0.$ We call such transformations {\it admissible}.  They act on  the  $a,b,s,d,f$ in (\ref{deform})
separately i.e.  $a,b,s,d,f$ are "`tensors"' with respect to the group of admissible transformations. General transformations (\ref{basis}) mix together these sets.

A priori we have two classes of solutions. The first class  is defined by $a^{33}_{12}=1$. We call such solutions {\it long}. For solutions from the second class we have $r^{33}_{12}=0$ after any transformation (\ref{basis}). We call them {\it short} solutions.
It is easy to verify that the short solutions are defined by the following linear relations:
\begin{equation}\begin{array}{c}\label{short}
r^{33}_{12}=r^{11}_{32}=r^{22}_{31}=0, \qquad r^{21}_{12}+r^{31}_{31}+r^{32}_{23}=r^{21}_{21}+r^{31}_{13}+r^{32}_{32},
\qquad r^{31}_{23}=r^{11}_{21}+r^{31}_{32}, \\[4mm] r^{32}_{31}=r^{22}_{21}+r^{32}_{13}, \qquad
 r^{31}_{12}=r^{31}_{21}+r^{33}_{32}, \qquad r^{32}_{21}=r^{32}_{12}+r^{33}_{31},  \\[4mm]
  r^{21}_{32}=r^{11}_{31}+r^{21}_{23}, \qquad r^{21}_{31}=r^{21}_{13}+r^{22}_{32}.
\end{array}
\end{equation}

{\bf Lemma 1.} Let $r$ be a solution of (\ref{yang}). If both $r$ and $T(r)$, where $T$ is defined by (\ref{sig}), are short, then $r=0$.

{\bf Scheme of a proof.} The  invariance of the conditions $r^{33}_{12}=r^{12}_{33}=0$ required gives us 20 linear relations between 36 independent components of $r$. Finding 20 components from the linear system and substituting them into (\ref{r1}), (\ref{r2}), we arrive at a big but simple system of quadratic equations, which contains many equations of the form $Z^2=0,$ where $Z$ is a linear expression in components of $r$. $\square$

In this section we consider long solutions.   Note that
since
\begin{equation}\label{sol0}
r^{33}_{21}=-r^{33}_{12}=1
\end{equation}
is a solution of (\ref{r1}), (\ref{r2}), the long solutions are deformations of (\ref{sol0}).

Equating  the coefficients at $\varepsilon$ in  (\ref{r2}) to zero, we get
$$
b^{31}_{22}=b^{32}_{11}=0, \qquad   b^{31}_{11}=b^{32}_{12}+b^{32}_{21}, \qquad b^{32}_{22}=b^{31}_{12}+b^{31}_{21}.
$$
Coefficients at $\varepsilon^2$ allow us to express 11 of 14 non-zero elements of the set $c$ in terms of  $b$.
Relations at $\varepsilon^3$ give us all 10 elements of $d$ and from $\varepsilon^4$ we find $f^{21}_{33}=0.$  The elements $b^{33}_{31}, b^{33}_{32},b^{32}_{21}, b^{31}_{12},b^{32}_{12}, b^{31}_{21}$ of the set $b$ and  the elements $c^{32}_{32}, c^{31}_{32}, c^{32}_{31}$ of $c$ remain to be undefined.

{\bf Lemma 2}. For any long solution $r$ the solution $T(r)$ is short.

{\bf Scheme of a proof.} Using the expressions for elements of $r$ in terms of the remaining 9 elements, we can easily verify that the relations (\ref{short}) for  $T(r)$ are fulfilled. $\square$.

It follows from Lemma 1 and Lemma 2 that the sets of long and short solutions are complementary with respect to the involution $T$.

Let us introduce three two-dimensional vectors
$$X=(b^{33}_{31}, b^{33}_{32} ), \qquad Y=(b^{32}_{21}, b^{31}_{12} ), \qquad Z=(b^{32}_{12}, b^{31}_{21} ).$$
The remaining equations from the system (\ref{r1}), (\ref{r2}) direct onto the following separation: \linebreak
{\bf Case 1:} $Z\ne -Y$ and  {\bf Case 2:} $Z=-Y$.

In Case 1 the  elements $c^{32}_{32}, c^{31}_{32}, c^{32}_{31}$  are determined and the whole system (\ref{r1}), (\ref{r2}) is equivalent to the only one relation
\begin{equation}\label{rel1}
x_2 y_1-x_1 y_2+x_2 z_1-x_1 z_2+y_2 z_1-y_2 z_2=0,
\end{equation}
where $x_i,y_i,z_i$ are corresponding components of $X,Y$ and $Z.$

In Case 2 the system (\ref{r1}), (\ref{r2}) is equivalent to the relation
\begin{equation}\label{rel2}
(x_2 y_1-c^{32}_{32})^2+(c^{31}_{32}+x_2 y_2)( c^{32}_{31}-x_1 y_1)=0.
\end{equation}

Now we should simplify the vectors $X,Y,Z$ by admissible transformations.  It is easy to see that the vectors  are transformed as follows:
$$X\to X Q,\qquad Y\to Y Q,\qquad Z\to Z Q,$$
where $Q$ is arbitrary non-degenerate matrix.

In Case 1 we have the following possibilities:
\begin{itemize}
\item {\bf Case 1-1}: $X$ is not parallel to $Y.$ In this case we reduce them to $X=(1,0), \, Y=(0,1)$. Then it follows from (\ref{rel1}) that $z_2=z_1-1.$
\item {\bf Case 1-2}: $X$ is not parallel to $Z$ and we arrive at  $X=(1,0), \, Z=(0,1)$ and $y_2=-y_1-1.$
\item {\bf Case 1-3}: $Y$ is not parallel to $Z$ and we obtain  $Y=(1,0), \, Z=(0,1)$ and $x_2=x_1+1.$
\item {\bf Case 1-4}: All three vectors are parallel: $X=(u,0),\, Y=(v,0),\, Z=(w,0),$ where   $v+w\ne 0$. Without loss of generality we put $w=1-v.$
\end{itemize}
The corresponding solutions of (\ref{r1}), (\ref{r2}) are given by:
\begin{equation}\begin{array}{c}\label{Case11}
r^{33}_{21}= r^{31}_{13}= r^{31}_{12} = r^{33}_{31}=1, \\[4mm]
r^{12}_{11}= r^{12}_{21}= r^{12}_{31}= r^{31}_{11}=r^{32}_{12}= r^{32}_{13}= r^{32}_{22}= r^{32}_{23}=r^{32}_{32}= r^{32}_{33}=u,  \\[4mm]
r^{11}_{12}= r^{11}_{13}= r^{31}_{21} = r^{31}_{31}=u-1;
\end{array}
\end{equation}
\phantom{y}
\begin{equation} \begin{array}{c}\label{Case12}
r^{33}_{21}= r^{31}_{31}= r^{31}_{21} = r^{33}_{31}=1, \\[4mm]
r^{12}_{11}= r^{21}_{21}= r^{21}_{31}= r^{23}_{22}=r^{23}_{23}= r^{23}_{32}= r^{23}_{33}= r^{31}_{11}=r^{32}_{21}= r^{32}_{31}=u ,  \\[4mm] r^{11}_{21}= r^{11}_{31}= r^{13}_{21} = r^{13}_{31}=u+1;
\end{array}
\end{equation}
\phantom{y}
\begin{equation}\begin{array}{c}\label{Case13}
r^{33}_{21}= r^{13}_{13}= r^{23}_{32}= r^{31}_{11} = r^{31}_{21}=r^{32}_{21}= r^{32}_{22} = r^{33}_{21}=1, \\[4mm]
r^{12}_{11}= r^{12}_{12}= r^{21}_{31}= r^{23}_{33}=r^{32}_{31}= r^{32}_{32}= r^{33}_{31}=u , \\[4mm]
r^{12}_{21}= r^{12}_{22}= r^{13}_{31}= r^{13}_{32}=r^{21}_{32}= r^{31}_{33}= r^{33}_{32}=u+1;
\end{array}
\end{equation}
and
\begin{equation}\begin{array}{c}\label{Case14}
r^{33}_{21}=r^{31}_{11}=1 \qquad r^{33}_{31}=r^{12}_{11}=u, \qquad  r^{32}_{21} =v, \qquad r^{32}_{12}=1-v,     \qquad r^{32}_{31}=u v,\\[4mm]
r^{32}_{13}=u (1-v), \qquad r^{22}_{12}=v (1-v),       \qquad r^{22}_{13}=u v (1-v).
\end{array}
\end{equation}
Other components of $r$ are defined by (\ref{r1}) or are equal to zero.

In Case 2 we have the following subcases:
\begin{itemize}
\item {\bf Case 2-1}: $X$ is not parallel to $Y$ and we  reduce them to $X=(1,0), \, Y=(0,1)$. Then (\ref{rel2}) can be parameterized as follows:  $c^{31}_{32}=u^2,$
$c^{32}_{31}=-v^2,$ and $c^{32}_{32}=u v,$
\item {\bf Case 2-2}: $X$ is parallel to $Y$ and we have $X=(\alpha,0), \, Y=(\beta,0)$,  $c^{31}_{32}=u^2,$ \linebreak
$c^{32}_{31}=-v^2+\alpha \beta,\,$  $\,c^{32}_{32}=u v.$
 \end{itemize}

They produce the following solutions:
\begin{equation}\begin{array}{c}\label{Case21}
r^{33}_{21}= r^{13}_{12}= r^{31}_{12} = r^{33}_{31}=1, \qquad r^{11}_{23}= r^{13}_{32} = r^{31}_{32}=u^2, \qquad   r^{12}_{12}= r^{21}_{12} = r^{23}_{32}=r^{32}_{32}=u v, \\[4mm]
r^{12}_{31}= r^{21}_{31}= r^{22}_{12}= r^{22}_{32}=r^{23}_{13}= r^{32}_{13}=v^2,  \\[4mm]
r^{11}_{12}=u^2-1, \qquad  r^{13}_{13}= r^{31}_{13} =u v+1, \qquad  r^{11}_{31}=2 u v+1,
\end{array}
\end{equation}
\phantom{y}
\begin{equation} \begin{array}{c}\label{Case22}
r^{33}_{21}=1, \qquad  r^{33}_{31}= p, \qquad r^{23}_{21} = r^{32}_{21}=q, \qquad r^{11}_{12}= r^{13}_{32} = r^{31}_{32}=u^2, \\[4mm]
r^{12}_{12}= r^{21}_{12}= r^{13}_{13}= r^{31}_{13}=r^{32}_{32}= r^{23}_{32}= u v,  \qquad r^{11}_{31}= r^{12}_{32} = r^{21}_{32}= q u^2,  \\[4mm] r^{23}_{31}= r^{32}_{31}=p q - v^2, \qquad  r^{22}_{32} = 2 q u v, \qquad  r^{22}_{31}=q (pq-v^2), \qquad  r^{22}_{21}=q^2-v^2.
\end{array}
\end{equation}

The corresponding short solutions can be found by applying the involution $T$ to (\ref{Case11})-(\ref{Case22}).
This completes the preliminary classification.

 \section{Equivalence}.

Now we should find  solutions  non-equivalent with respect to the whole group of transformations (\ref{basis}). It will be shown that the parameters in all families of solutions found in Section 2 are inessential. This means that each family provides a solution corresponding to generic values of the parameters and several special solutions.

To bring a solution to a simple form we consider  matrices $M$ and $N$ with entries $M^i_j=r^{\alpha i}_{\alpha j}$ and $N^i_j=r^{\alpha i}_{j \alpha}.$  Under (\ref{basis}) these matrices are transformed just as $M\to G^{-1} M G$ and $N\to G^{-1} N G$. So invariants of the pair $(M,N)$ (see \cite{CProc}) are also invariants for the corresponding solution $r$. In particular, all coefficients of the characteristic polynomial $S=Det({\bf 1}+\lambda M+\mu N)$ are invariants of $r.$
It can be straightforwardly checked that $S\equiv 1$ for each solution from Section 2. This implies that all eigen-values for any linear combination $M$ and $N$  are equal to zero.

All solutions $r$ can be subdivided into the following three groups: Group 1 consists of solutions for which the Jordan form of   $Q=\lambda M+\mu N$ for generic $\lambda,\mu$ is the big Jordan block; for Group 2 the Jordan form is the small Jordan block; and $M=N=0$ for Group 3.
 For equations of Group 1 the Jordan basis for $Q$ is a basis in which the components of $r$ become very simple.

 Consider the family of solutions (\ref{Case22}) from Case 2-2. The generic solution corresponding to  $u (p+q)\ne 0$ belongs to Group 1. It is equivalent to
\begin{equation} \label{sol1}
r^{11}_{12}=r^{11}_{32}=r^{12}_{13}=r^{21}_{13}= r^{22}_{23}=1.
\end{equation}

Three non-equivalent solutions of Group 2 correspond  to  {\bf a}: $q=-p, u\ne 0$; {\bf b}:  $u=v=0, \, p+q\ne 0$ and {\bf c}: $u=0,\, v=\frac{p+q}{2}, \, p+q\ne 0.$ They are equivalent to
\begin{equation} \label{sol2}
r^{11}_{32}=r^{12}_{13}=r^{21}_{13}=r^{22}_{23}=1.
\end{equation}
\begin{equation} \label{sol3}
r^{22}_{23}= 1.
\end{equation}
and
\begin{equation} \label{sol4}
r^{12}_{13}=r^{21}_{13}=r^{22}_{23}=1.
\end{equation}
The  only one non-equivalent solution  of Group 3 is given by (\ref{sol0}). It corresponds to $u=v=p=q=0$.

It turns out that any solution of Class 2-1 is equivalent to one of   (\ref{sol1})- (\ref{sol4}).

Case 1-1 with $u\ne 0,\, u\ne 1$ and with $u=1$ produces two new solutions equivalent to:
\begin{equation} \label{sol5}
r^{11}_{13}=r^{12}_{13}=r^{32}_{33}=1.
\end{equation}
and
\begin{equation} \label{sol6}
r^{12}_{31}=r^{32}_{33}=1.
\end{equation}

At last, Case 1-2 with $u=-1$ gives us a solution equivalent to
\begin{equation} \label{sol7}
r^{12}_{13}=r^{32}_{33}=1.
\end{equation}

\subsection{Final result}

Our previous computations lead to the following

{\bf Theorem 1.} Any long solution $r$ of the system (\ref{r1}), (\ref{r2}) can be reduced to one of non-equivalent equations (\ref{sol0}), (\ref{sol1})- (\ref{sol7})  by a transformation (\ref{basis}). Equation (\ref{sol1}) belongs to Group 1, equations (\ref{sol2})- (\ref{sol7}) form Group 2 and equation (\ref{sol0})  belongs to Group 3. To obtain all short solutions we have to apply the involution  $T$ defined by (\ref{sig}) to the long solutions described above. $\square$

Thus we proved that there exist 16 solutions non-equivalent with respect to the group of transformations (\ref{basis}).

Two solutions $r$ and  $T(r)$ of Group 1, where $r$ is defined by (\ref{sol1}),  correspond to anti-Frobenius associative algebras of $3\times 3$-matrices with one zero row and one zero column, correspondingly.
It would be interesting to investigate the anti-Frobenius associative algebras corresponding to other solutions. It could give a hint how to formulate a reasonable classification problem for anti-Frobenius associative algebras with arbitrary $n$. Some classification of anti-Frobenius Lie algebras can be found in \cite{elash}.

\vskip.3cm
\noindent
{\bf Acknowledgments.}
The author is grateful to A. Odesskii, V. Roubtsov, A. Zobnin
for useful discussions.
The research was  partially supported by the RFBR grant 11-01-00341-a.


\bigskip

\begin{thebibliography}{10}

\bibitem{Rota} G.-C. Rota,
\newblock{Baxter operators, an introduction. Gian-Carlo Rota on combinatorics},
\newblock{\em Contemp. Math.}, Birkhauser Boston, Boston MA,  {\bf 57}(6), 504--512, 1995.

\bibitem{Sch}
 T.~Schedler,
\newblock{ Poisson algebras and Yang-Baxter equations. Advances in quantum computation}, 91ï¿-106, \newblock{\em Contemp. Math.}, {\bf 482}, Amer. Math. Soc., Providence, RI, 2009.,

\bibitem{MikSok}  A.V.~Mikhailov  and V.V.~Sokolov,
\newblock{ Integrable ODEs on Associative Algebras}.
\newblock{\em Comm. Math. Phys}. {\bf 211}, 231-251, 2000.

\bibitem{OdRubSok1}
 A.V.~Odesskii, V.N.~Rubtsov and V.V.~Sokolov,
\newblock{Bi-hamiltonian  Ordinary Differential Equations with Matrix Coefficients},
\newblock{\em  Theor. Math. Phys.}, {\bf 171}, 442--447,  2012.

\bibitem{VdB}
 M.~Van den Bergh,
\newblock{Double Poisson algebras},
\newblock{\em Trans. Amer. Math. Soc.}, {\bf 360}, no. 11, 5711ï¿-5769, 2008.

\bibitem{OdRubSok2}
 A.V.~Odesskii, V.N.~Rubtsov and V. V.~Sokolov,
\newblock{Double Poisson
brackets on free associative algebras},
\newblock{\em arXiv:1208.2935}, 2012.

\bibitem{Agui} M. Aguiar,
\newblock{ On the associative analog of Lie bialgebras.}
 \newblock{\em J. Algebra}, {\bf 244}, 492--532, 2001.


\bibitem{CProc}
 C.~Procesi,
\newblock{The invariant theory of $n\times n$ matrices},
\newblock{\em  Advances in  Math.}, {\bf 19 }, 306--381,  1976.

\bibitem{elash}  A. G.~Elashvili,
\newblock{Frobenius Lie algebras.}
\newblock{\em Functional Anal. Appl.}, {\bf 16}, no. 4, 326--328, 1983.


\end{thebibliography}
\end{document}